# Identifying Requirements Affecting Latency in a Softwarized Network for Future 5G and Beyond

Idris Badmus, Abdelquoddouss Laghrissi, Marja Matinmikko-Blue and Ari Pouttu
Centre for Wireless Communication
University of Oulu
Oulu, Finland
firstname.lastname@oulu.fi

*Abstract*—The concept of a softwarized network leveraging technologies such as SDN/NFV, comes with different merits such as decreased Operational Expenses (OPEX), and less dependency on underlying hardware components. With the amount of increased flexibility, reconfigurability and programmability attributed to future technologies (i.e., 5G and beyond), and towards the complete network virtualization and softwarization, a new set of requirements/parameters can be identified affecting the latency in a virtualized network. In this paper, we identify different latency requirements for a virtualized network. These requirements include the Virtual Network Function (VNF) deployment time, establishment/connection time and application instantiation time. We further test how some factors such as VNFs' resource usage, the applications running within the VNF and the shared status of the VNF, coordinately affect the identified latency requirement for a virtualized network. Experimentally, for performance analysis, we deploy a softwarized network based on the ETSI-NFV architecture, using open source tools. The results show that the new set of latency requirements is relevant for consideration in order to achieve an overall ultra-reliable low latency and how different the factors can affect these new requirements, especially in the core network. Furthermore, the result of our performance analysis proves the trade-off between latency of a virtualized network and the resource usage of the VNFs.

## I. INTRODUCTION

The idea of having an end-to-end softwarized or virtualized network has been proposed as a key requirement for achieving ultra-reliable low latency in future technologies of 5G and beyond [1]–[3]. The concept of network softwarization involves changing the paradigm of network deployment from the existing commercial off-the-shelf hardware type to a more software-dependent and programmable network [4], [5]. To achieve this, different network components in the access and core networks are decoupled from the static underlying vendor specific component, and to software services and applications that can perform the same functionalities efficiently and which are less vendor-dependent [1], [6]. To achieve network virtualization and softwarization, it is vital to leverage technologies such as Software Defined Network and Network Function Virtualization (SDN/NFV) [7], [8]. These technologies serve as enablers in achieving a fully virtualized network. SDN deals with the separation of the control plane from the data plane, and the application of different rules and policies for forwarding traffic from one router or switch to another using its centralized logical controller, called the SDN controller [7]. NFV on the other hand serves as a major entity in the 5G core for the support of visualization [6], [8]. NFV reduces extensively the dependencies on hardware and paves the way for a flexible deployment of end-to-end networks with an eased interaction between the core network and the RAN [9].

Generally, the concept of network softwarization or virtualization leveraging SDN/NFV aims to achieve different objectives in future network technologies; 5G and beyond. These include less dependencies on existing vendors off-the-shelf hardware solutions [7], reduced OPEX and increased CAPEX[10]. Also, one other important aim of a virtualized network is to extensively reduce the latency, especially at the core network [11]. However, it is paramount to understand that while these technologies are fulfilling some latency requirements with respect to future networks [12], a new set of latency requirements emerges due to the deployment needs. If these set of requirements are not properly addressed, their impact might forfeit the initial aim of achieving ultra-reliable low latency in a virtualized network.

In this paper, new latency requirements with regards to a virtualized network are identified. These requirements include the VNF deployment time, the connection/establishment time of VNFs, and the application instantiation time. These three latency requirements are only specific to a virtualized network whereby network components are deployed as VNFs. Basically, unlike the conventional latency requirement in the core network such as the placement of VNFs [13] and modifying the core network architecture by introducing SDN controllers to handle specific components [14], the latency requirements introduced in this paper are peculiar to any type of VNF deployment. Furthermore, since VNFs form the basis of a virtualized 5G network, the requirements will ultimately affect the overall latency in a virtualized network. Hence, it is crucial to identify these requirements, test the factors that can affect them, and conclude on how they can be reduced efficiently or completely eliminated.

We further experiment the factors contributing to the identified latency requirements in a virtualized network. These factors include the available resources/resource usage, the number of application services running in the VNF and the shared status of the VNFs (i.e., whether the VNFs are deployed as shared or non-shared). To measure the performance analysis of these factors, we deploy a virtualized network environment based on ETSI-NFV architecture [15] using open source tools such as Open Source MANO (OSM) and OpenStack cloud infrastructure.

The remaining part of the paper is arranged as follows. Section II discusses the literature review regarding latency in a virtualized network. The identified latency requirements are described in section III, and the experimental results of the factors affecting the latency requirements are described in section IV. The concluding section contains the summary of the work done so far and possible future contributions.

## II. LITERATURE REVIEW

The idea of latency reduction in the core network has been one of the main purposes of the SDN/NFV technologies through the separation of the control plane from the data plane [7], [11]. For 4G networks, different proposals and techniques have been introduced in addressing latency concerns in a virtualized core network. Authors in [13] discussed the concept of VNF placement, where network functionalities within the virtualized core network react to latency when they are placed at different locations in terms of data centers or cloud infrastructures. These concepts extend the idea of deploying the network across multiple domains [16]. Another approach to addressing latency within a virtualized Evolved Packet Core (EPC) was proposed by authors in [14], [17], in order to reduce the number of Server Gateways (SGWs). Even though this approach is still in relation with VNF placement, the reduction in the number of SGWs paves the way for an SDN controller implementation of the SGW. Since the SGW is responsible for session creation, the SDN controller will enable an intelligent routing of packets and increase quality of service (QoS) handling. With the implementation of this approach, latency within the core network was reduced to some extent. The Authors in [18] proposed the idea of visualizing the EPC as virtual machines in different locations by introducing a carrier cloud architecture. Using he Follow-me Cloud concept, all components on the virtualized EPC are managed by the carrier cloud, and each component can keep track of the mobility of users which will reduce latency in the core especially during handovers. Furthermore, authors in [19] approaches latency in a virtualized network with regards to the delay at the MME. They proposed the idea of having a logical SDN-based mobility management application (MMA) and studied how the internal mechanism of the MMA and the SDN controller can be coordinately used in reducing the end-to-end latency.

To the best knowledge of the authors, no previous works had discussed the impact of factors such as resource usage, the number of applications running in VNFs, and the VNFs sharing status on the identified latency requirements in a virtual network. As such, this paper identifies this new set of requirements, and experiments their impact on latency.

## III. IDENTIFIED LATENCY REQUIREMENT IN A SOFTWARIZED NETWORK

With the virtualized networks being introduced in previous works [14], [17]–[19], three latency requirements are identified in this paper for a virtualized network, highlighting factors that affect these requirements, and which are tested in a deployed virtualized network environment. The identified latency requirements include the VNF deployment time or instantiation time, connection time or establishment time between different VNFs, and the application service instantiation time.

The **VNF deployment time** describes the instantiation time of the network service. This covers the time when a VNF is deployed as a network service to the time the VNF starts operating. Different factors can affect the deployment time and it can lead to an overall increased latency in the network. Since VNFs will be instantiated at a required time for a service specific functionality, it is important they are only deployed at that time to avoid the usage excess of available resources. Hence, one of the main factors that affect the deployment time is the flavor of the VNF's Virtual Descriptor Unit (VDU), and mostly the available resources in the Virtual Infrastructure Machine (VIM) or the cloud infrastructure where the Network Function Virtualization Infrastructure (NFVI) is running. These resources include computing, networking and storage resources. Generally, there is always a tradeoff between latency and available resources, so if there are enough resources in the network, the latency due to deployment time will be minimized extensively. However, due to the OPEX of running a virtualized network, network operators wants to maximize resources to cover multiple network services running different VNFs. With technologies such as network slicing which is completely built on a virtualized infrastructure [20], each network slice instance requires a new set of VNF instantiation, and hence it is vital to be able to manage the available resources and still achieve minimal latency due to VNF instantiation.

The **connection time or establishment time** is the period within the VNF deployment when all network services responsible for a network deployment are connected together, and all interfaces are open for connection. This is also required in e network slicing whereby all Network Slice Subnet Instances (NSSIs) from Access, Core and Transport Network [21], are connected or established together using the Network Slice Template (NST) to get the required Network Slice Instance (NSI). Factors affecting the connection time latency requirement in a virtualized network include the available resources and shared status of the VNF.

The **application instantiation time** is the period when different application services responsible for network functionalities are up and running in the VNF. Generally, in a virtualized network, different network components or network functionalities (e.g., HSS, MME, PGW and SGW) are deployed as application services running within a VNF. These running applications can range from the regular EPC/Radio network functionalities to general application services (e.g., video processing or database manager). These functions are deployed as application services and they consume some amount of resources in the network. The instantiation of these application services is different from the deployment time of VNFs; the application services are deployed after the VNF is up and running. Thus, factors affecting the application instantiation time of a VNF, which can add to the overall latency in a virtualized network include the following; The number of applications that are running in the VNF, the VDU flavor for the VNF, which include the allocated computing, storage and network capabilities attributed to the specific VNFs, and the functionality or service status of the application (i.e., either the application is always up and running or the application is only up when needed). Thus, to maximize OPEX, the number of resources to be used by the application in a specific VNF can be scaled down or up from the orchestrator when the VNF is deployed which can affect the overall latency of the VNF especially when the application service is of vital use.

In summary, the identified set of requirements can affect the overall latency of a network and it is important to understand how it can contribute to latency. In this vein, we prove the impact of these parameters on the overall latency of a virtualized or softwarized network.

## IV. EXPERIMENTAL RESULTS OF FACTORS AFFECTING THE IDENTIFIED LATENCY REQUIREMENTS

In this section, we experiment and test the impact of three important factors affecting the identified latency requirements in a virtualized network. These include the VNF memory utilization or the available memory, the applications running in the VNF, and the shared status of the VNF. Each of these factors coordinately contribute to the identified latency requirements for a virtualized network in one way or the other. To create the experiment for testing these factors, we deployed a virtualized network environment using open source tools. We used OSM for the management and orchestration of the VNFs, OpenStack cloud infrastructure for the Machine VIM and for implementing NFVI layer. In the experiment we also used some dummy Ubuntu VNF images of different sizes and flavors. As for the virtual EPC, we used the NextEPC solution. We integrated Gnocchi and Ceilometer in our OpenStack cloud to expose the performance metrics APIs of each deployed VNF and then Prometheus for displaying the exposed metrics in a graphical form. Each of the factors are tested experimentally, and their impact on the identified latency requirements is evaluated as follows.

### A. Impact of memory utilization

This factor describes the available memory in the VIM after a VNF is deployed and how this can contribute to the latency during the deployment of the VNF. Basically, for a VNF to be deployed as a network service function, the declared flavor of the VNF (i.e. properties of the VDU need to be available in the VIM). One of these properties is the memory to be allocated for the VNF. This memory must be made available in the VIM in order for the VNF to be deployed. This will determine the memory utilization and it greatly influences the latency of deploying the VNF. Since the memory in the VIM server is made available whenever a VNF service is to be deployed. Thus, when the VNF is instantiated, if there is an excess in available memory in this VIM, the VNF can be deployed easily, but if there aren't enough memory or if the VIM is low on memory, the server tries to scavenge memory from other services and hence this affects the time of deployment which ultimately affects the latency during deployment. For instance, when a VNF with VDU of 2GB memory is deployed, if the VIM has more than 32GB memory available, the VNF can be deployed easily with nearly zero deployment latency, however this is not the case in most virtualized network deployments since there would be multiple deployments happening and the network operator is trying to maximize resources, hence the VIM will try to scavenge memory to deploy the VNF. Since there are multiple VNFs instantiating at a single time to form a network service, the search for memory will contribute to the overall latency of the network. In our experiment, we prove the impact of the available memory on the overall deployment time of a VNF. Using a deployed virtualized network environment, we use a dummy Ubuntu VNF of image size 2GB and a VDU flavor of 2 CPUs, 2048 MB of Memory and one network connection point interface. Using OSM, we orchestrate the VNF deployment as a network service using a VIM server of 16 GB memory (with less than 8GB available), Intel core i7 running at 2.6GHz. We carry out 5 instantiations of the same VNF from OSM, and we measure the available resources using Linux shell command *free –h*, and create a trail time history of the deployment in seconds against the available memory until the VNF is instantiated. The data containing available memory at different times in seconds during the deployment of the VNF is depicted in Fig.1.

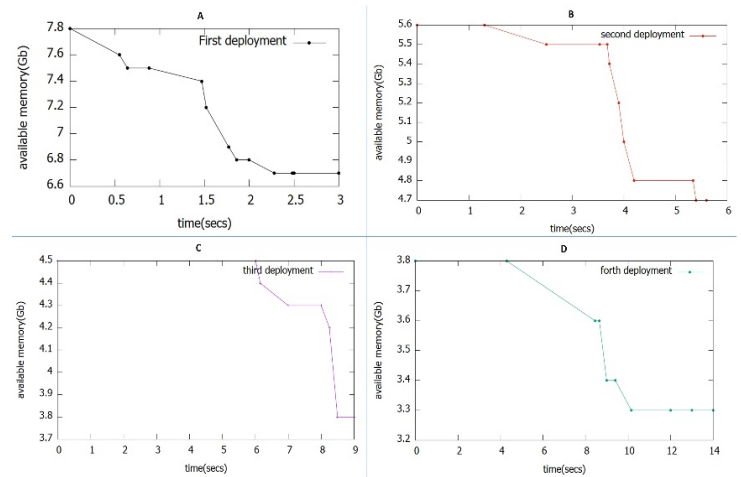

Fig. 1. Graph of available memory against time for first four deployments.

The result of the first four deployments can be seen in Fig. 1. It can be seen that as the amount of memory is reducing with each deployed VNF, the deployment time is longer even though the needed memory for deployment is still available. Hence, the deployment time is longer since the server needs to scavenge for memory. After the VIM server is reduced to less than a 3GB, and the fifth deployment is made the VIM can no longer deploy the VNF since it is out of memory and after a longer deployment time, the VNF instantiation fails and the output can be seen in Fig 2. Thus, in general, if the size of available memory is large, the VNF deployment time will be minimized and any latency from this part can be overall avoided. However, when the memory is getting low or when there are multiple deployments, there will be a deployment time delay in the network.

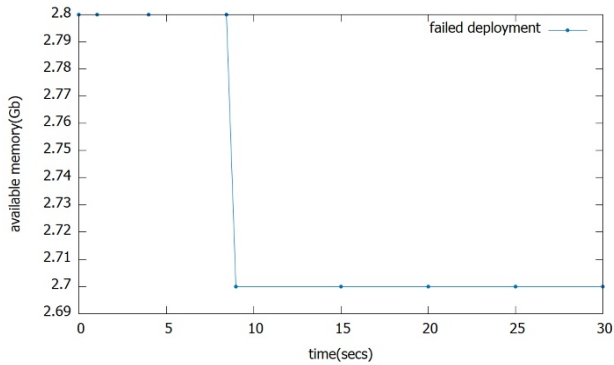

Fig. 2. Graph showing the failed deployment after there isn't enough memory.

### B. Impact of running application service in the VNF

To test the impact of running applications on the overall latency of a virtualized network, we experiment the latency of a VNF running no applications and a VNF running some applications. The impact of running applications in a VNF can affect the deployment time and the application instantiation time. Generally, some applications are always up and running in the VNF, so whenever the VNF is deployed, they are automatically deployed and they are up and running while some applications are tuned on while the VNF is already running. This set of applications consumes resources and contributes to the overall latency in the network.

To create a VNF running an application service, we deployed a Virtual EPC using the NextEPC solution. The NextEPC contains an all-in-one virtualized EPC solution that will be triggered when the application is needed. Thus since it is an open source solution, we were able to perform some operations on the EPC without jeopardizing the EPC application service itself. We separated the network components in the cloud by shutting down other network functions and allowing the interface IP address to be passed when deploying the EPC. For this experiment, we combined the application service responsible for SGW, PGW and MME into a single VNF and it is made to be always up and running whenever the VNF is instantiated to reduce the connection time to other network components' VNFs.

In our experiment, we deploy the VNF running the SGW, PGW and MME as an application VNF against a dummy Ubuntu image, both deployed with the same VDU flavor. We compare the overall instantiation time which in this case will include the deployment time and the application instantiation time of the dummy VNF and VNF running SPGWMME application. As seen in Fig. 3, it can be observed that despite running on the same VDU flavor, the overall instantiation time of the VNF running the application is much more than the time for the VNF running no applications. Hence, the amount of applications and the type of applications running in a VNF contribute to the overall latency of the VNF. Since it is impossible to completely remove all application services from running within a VNF, we can decrease the number of applications running or we can move critical applications to the edge to reduce latency.

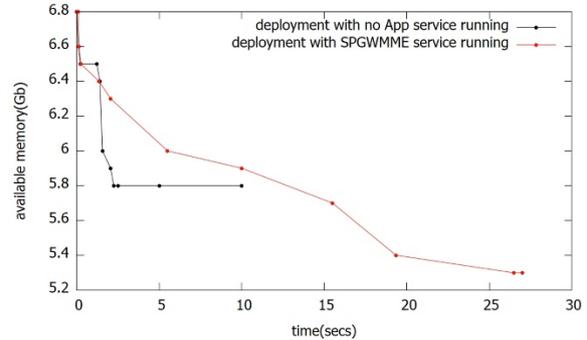

Fig. 3. Graph showing the impact of application running in a VNF.

### C. Impact of shared and non-shared VNF

Lastly in this experiment, the impact of shared and non-shared resources is tested for the implementation of VNFs. Basically, 3GPP [22] introduced the concept of shared and non-shared constituents in network slicing, where a VNF or a set of VNFs deployed as an NSSI can be shared by different network slice instances. The sharing of a VNF can definitely affect the overall latency of the network considering requirements such as the establishment or connection time, and deployment time of the VNF. Authors in [21], [23] show how shared and non-shared constituents of a network slice can be used to categorize network slicing in a micro-operator. In the same fashion, we show the impact of shared and non-shared VNFs on the overall deployment time and establishment time of a VNF.

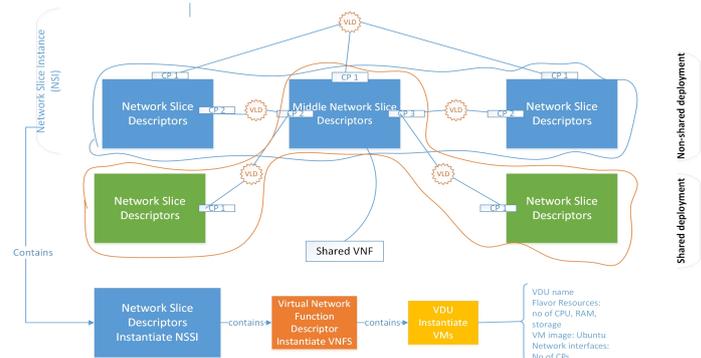

Fig. 4. Shared and non-shared deployment arrangement of NSSIs.

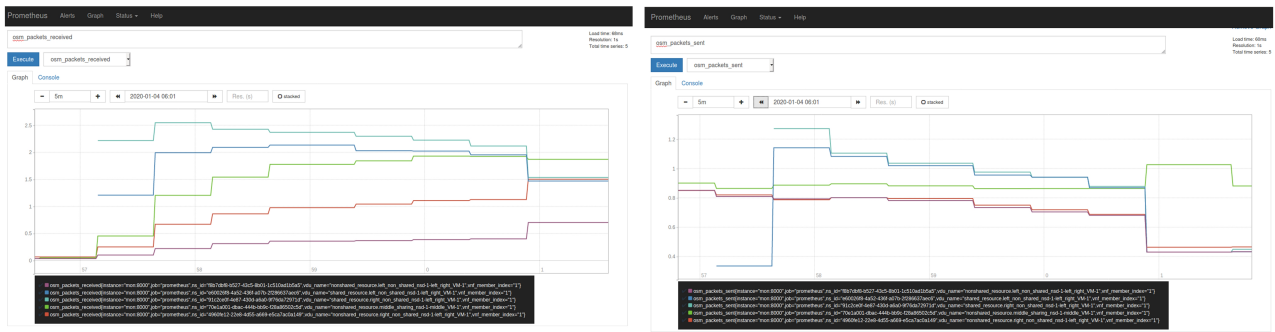

Fig. 5. Prometheus output for shared and non-shared deployment showing packet received and packet sent by VNFs.

To achieve this, we created a network slicing solution to test the shared VNFs by establishing a shared and non-shared deployed NSIs containing 3 NSSIs; each NSSI containing one Ubuntu VNF. The shared and non-shared NSIs are arranged as seen in Fig. 4, to form the network slice. Using the NST in OSM, we share the middle VNF between the two different NSIs. This means that we first deploy the non-shared NSI constituting of 3 NSSIs whereby each NSSI constitutes one VNF as the middle VNF is made sharable. Then, we deploy the second shared NSI with the first and last NSSI, and the middle VNF is shared with the first deployment. For this particular experiment, the VDU image of the VNFs are of size 400MB and as such, the VDU flavor contains 1024Mb of memory, and 2 CPUs allocated. From the graph in Fig. 6, it can be seen that even though the non-shared deployment includes 3 VNFs, the total deployment and connection time is less than that of the shared VNFs with just 2 VNFs. This is due to the fact that the shared deployment involves an establishment of connection with the already deployed middle VNF which takes time and hence the total time is extended.

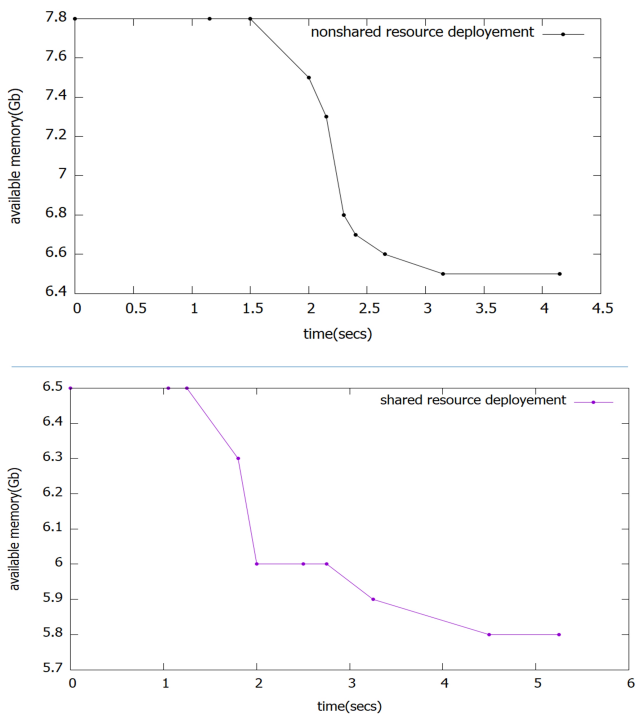

Fig. 6. Graph showing impact of shared and non-shared deployment.

This is expected to be longer if there are applications running in the VNFs. Meanwhile, from the memory perspective, it can be seen that the memory usage of the non-shared deployment is much more than the memory usage for the shared deployment and this is due to the number of VNFs in each deployment, hence the middle VNF contributed more memory usage to the non-shared deployment. We further check the impact of other factors such as the number of packets sent and received by the VNF during these deployments. To achieve this, we expose the performance metrics API from the VIM cloud infrastructure, using Gnocchi and Ceilometer, and we graphically viewed the exposed metrics in Prometheus. Fig. 5 shows the number of packets received and sent by each VNF. If we focus on the shared middle VNF which is represented in green, it can be seen that the number of packets received kept on receiving after the new shared deployment is activated. This is because the new shared deployment send more packets to the middle shared VNF to establish connection and hence it receives more packets, however the packet sent from the middle shared VNF remained on the same level despite the new deployment. The period for establishing this connection also contributes to the total latency in the shared deployment as seen in Fig. 6.

So basically to reduce latency, the non-shared deployment is better since there won't be any need to share and reconnect a new VNF every time. However, if we want to better manage resources, the shared deployment is better since it consumes less memory. This concept describes the tradeoff between the latency and resource usage in a virtualized network.

## V. CONCLUSION

This paper identifies a new set of requirements that can affect latency in a virtualized network. Even though these requirements are common to all VNF deployments, they can affect the overall ultra-reliable low latency required for 5G and beyond. Furthermore, the paper proves the impact of specific factors on the identified latency requirements. The factors include the available resources, the number of applications running in the VNF, and whether the VNFs are shared or not. The results from this paper confirm how these factors affect the latency. The tradeoff between the available resources for a VNF and the latency is also established. Future work will focus on identifying how to balance this trade-off and propose new solutions to reduce the general instantiation time of VNFs towards near-zero latency for technologies beyond 5G.

ACKNOWLEDGMENT

This work was partially supported by the European Union's Horizon 2020 Research and Innovation Program under the 5G!Drones project (Grant No. 857031), the MOSSAF Project, and the Academy of Finland 6Genesis project (Grant No. 318927).